\documentclass[journal]{IEEEtran}
\usepackage[utf8]{inputenc}

\usepackage{amsmath,amssymb}

\usepackage{color}
\usepackage{graphicx}

\usepackage{balance}
\usepackage{cite}

\newcommand{\Eb}{$E_\text{b}/N_0$}

\begin{document}
\title{{{Sensitivity Gains by Mismatched Probabilistic Shaping for Optical Communication Systems}}}
%
%
%

\author{Tobias~Fehenberger,~\IEEEmembership{Student Member,~IEEE}, Domani\c{c} Lavery,~\IEEEmembership{Member,~IEEE}, Robert Maher,~\IEEEmembership{Senior Member,~IEEE}, Alex Alvarado,~\IEEEmembership{Senior Member,~IEEE}, Polina Bayvel,~\IEEEmembership{Fellow,~IEEE}, and~Norbert~Hanik,~\IEEEmembership{Senior Member,~IEEE}
\thanks{Research supported by the Engineering and Physical Sciences Research Council (EPSRC) project UNLOC (EP/J017582/1), United Kingdom.}%
\thanks{Tobias Fehenberger and Norbert Hanik are with the Institute for Communications Engineering, Technical University of Munich (TUM), 80333 Munich, Germany (\mbox{Emails:}~tobias.fehenberger@tum.de, norbert.hanik@tum.de).}%
\thanks{Domani\c{c} Lavery, Robert Maher, Alex Alvarado, and Polina Bayvel are with the Optical Networks Group, University College London (UCL), London, WC1E 7JE, UK (\mbox{Emails:}~d.lavery@ee.ucl.ac.uk, r.maher@ucl.ac.uk, a.alvarado@ucl.ac.uk, p.bayvel@ucl.ac.uk).}
}%

\maketitle

\begin{abstract}
	Probabilistic shaping of quadrature amplitude modulation (QAM) is used to enhance the sensitivity of an optical communication system. Sensitivity gains of 0.43~dB and 0.8~dB are demonstrated in back-to-back experiments by shaping of 16QAM and 64QAM, respectively. Further, numerical simulations are used to prove the robustness of probabilistic shaping to a mismatch between the constellation used and the signal-to-noise ratio (SNR) of the channel. It is found that, accepting a 0.1~dB SNR penalty, only four shaping distributions are required to support these gains for 64QAM.	
\end{abstract}


\IEEEpeerreviewmaketitle

\section{Introduction}
	\IEEEPARstart{O}{ptical} fiber communication systems have experienced a dramatic evolution over the past two decades as the traffic demand has continued to grow \cite{CapacityCrunch}. In order to satisfy this demand for capacity, there has been a focus on increasing spectral efficiency by, e.g., reducing the guard interval in wavelength division multiplexing or employing high-order modulation formats. Square quadrature amplitude modulation (QAM) formats are the preferred alternative due to their simplicity in generation and detection as each constellation point is located on a regular square grid.

	A way to further increase spectral efficiency is by means of signal shaping \cite{FischerBook}.  In particular, geometrical shaping could be used, where constellation points are nonuniformly arranged on the complex plane.  Alternatively, symbols can be transmitted using nonuniform probabilities, which is known as \emph{probabilistic shaping}. For the additive white Gaussian noise (AWGN) channel with an average power constraint, both shaping techniques yield a sensitivity gain of up to $1.53$~dB relative to uniform QAM as the constellation cardinality tends to infinity {\cite[Sec.~VIII-A]{Wachsmann1999}}. This enhanced signal-to-noise ratio (SNR) will result in increased achievable information rate (AIR) for SNR-limited digital coherent transceivers \cite{Rob2015ECOC}. Therefore, constellation shaping is a promising candidate to improve spectral efficiencies in future fiber optic systems.

	Geometrical shaping has been demonstrated in fiber experiments \cite{GeoShapingExp1,GeoShapingExp2}, however, this shaping alternative poses stronger requirements on the effective number of bits (ENOB) of the digital-to-analog converter (DAC) due to the unequally spaced constellation points.  On the other hand, probabilistic shaping is a well-studied area of communication theory, see, e.g., \cite{forney1984efficient,FischerBook,Kschischang1993}, and a detailed review in \cite[Sec.~II]{BoechererLongShaping}. Probabilistic shaping is also known to give larger gains than geometrical shaping for square QAM, as shown in \cite[Fig.~4.8 (bottom)]{AlexBook}.

	{For fiber transmission in the presence of nonlinearities, probabilistic shaping has been shown to allow increased transmission performance. In split-step simulations, different input probability mass functions (PMFs) have been studied, such as a dyadic PMF as well as a PMF that is the result of  the Blahut-Arimoto algorithm \cite[Sec.~II]{MetodiManyToOneMapping}, ring-like constellations \cite[Sec.~III]{SmithFrankTrellisShaping}, and a Maxwell-Boltzmann PMF \cite[Sec.~V-B]{BeygiShellMappingShaping}, \cite[Sec.~2.3]{myProbShapingOFC}, \cite[Sec.~3.5]{FehenbergerOE2015}. Very recently, reach increases were experimentally demonstrated in \cite{ALUPDP2015} using a finite number of Maxwell-Boltzmann distributions. These distributions were chosen to target different net data rates, and shaping gains over a transmission range of more than 4500~km were reported.} 

	{In this Letter, we study the benefits of using probabilistically shaped QAM over uniform QAM. Both a back-to-back (B2B) single-carrier optical system and numerical simulations are considered. The contributions of the paper are twofold. First, a significant gain in SNR is demonstrated for a state-of-the-art digital coherent transceiver. To the best of our knowledge, this is the first experimental study of B2B transceiver sensitivity improvements due to probabilistic shaping in optical communications. Second, we investigate the effect of mismatched probabilistic shaping, i.e., how a mismatch between channel SNR and the SNR for which the input distribution is optimized affects shaping gains. We find that four input distributions are sufficient to obtain large shaping gains for 64QAM over a wide SNR range. To the best of our knowledge, this is the first numerical study of shaping robustness to a channel mismatch.}

\section{Probabilistic Shaping Method}\label{sec:shaping}
	Finding a good nonuniform input distribution for QAM constellations is a two-dimensional (2D) optimization problem. However, because of the symmetry of the 2D constellation, of the binary reflected Gray labeling which is typically used, and the AWGN channel for which the optimization is carried out, it is sufficient to consider a one-dimensional (1D) constellation. Under these assumptions, $M^2$-QAM constellations can be decomposed into a product of two constituent 1D (pulse amplitude modulation) constellations, each with $M$ constellation points. Without loss of generality, we therefore consider only one of the quadratures for the shaping optimization, however, we emphasize that all analysis in Sec.~\ref{sec:results} is performed for 2D ($M^2$-QAM) constellations on each polarization.

	Let $\boldsymbol{x}=[x_1, x_2,\dots,x_{M}]$ denote the real-valued constellation symbols, represented by the random variable $X$. We assume that the symbols are distributed according to a PMF $\boldsymbol{P}_X=[P_{X}(x_1), P_{X}(x_2),\dots,P_{X}(x_{M})]$ and that they are sorted in ascending order (i.e., $x_i\!<\!x_{i+1}$, $i=1,2,\ldots,M-1$). 

	{For shaping the input, we use a PMF from the family of Maxwell-Boltzmann distributions, which are well-known for the AWGN channel, see, e.g., \cite[Sec.~IV]{Kschischang1993}, \cite[Sec.~VIII-A]{Wachsmann1999}, and \cite[Sec.~III-C]{BoechererLongShaping}. Following the approach of \cite[Sec.~III-C]{BoechererLongShaping}, the shaped input is distributed as}
	\begin{align}\label{eq:shaping}
		P_X(x_i)=\frac{1}{\sum_{k=1}^M e^{-\nu x_k^2}}e^{-\nu x_i ^2},
	\end{align}
	where $\nu$ is a scaling factor. To find the optimal PMF among all distributions given by \eqref{eq:shaping}, let the positive scalar $\Delta$ denote a constellation scaling of $X$. Fixing $\Delta$ and the SNR for which the optimization is carried out (denoted \textit{shaping SNR} in the following), the scaling $\nu$ is chosen such that $E[|\Delta X|^2]=\text{SNR}={E_\text{s}}/{N_0}$ where $E_\text{s}$ represents the signal power and $N_0$ represents the noise power. The mutual information (MI) between the scaled channel input $\Delta X$ and the AWGN channel output is unimodal in $\Delta$, and thus, we choose the scaling factor that maximizes the MI. Although this optimization is carried out for symbol-wise MI, the loss from considering a bit-wise decoder \cite{AlexBook} gives a negligible AIR penalty \cite[Table III]{BoechererLongShaping}.

	Note that the input PMF in \eqref{eq:shaping} is known to be suboptimal for the AWGN channel, however, it has been shown to give near-capacity results \cite[Table I]{BoechererLongShaping}. Further, note that in order to choose the input PMF, the channel SNR must be known \emph{a priori} (see \eqref{eq:shaping}). We will show in Sec.~\ref{ssec:robustness} that imperfect SNR knowledge causes only a small penalty in comparison to a perfectly matched channel SNR and input PMF. 

\section{Experimental Setup}		
	\begin{figure}[t!]
	\footnotesize{
    \includegraphics{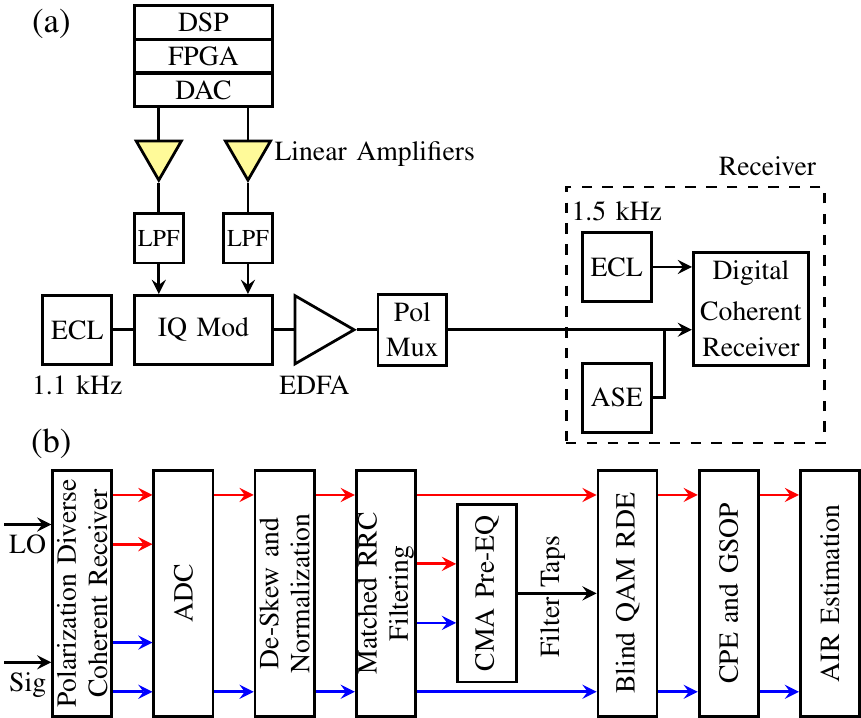}
	}
	\vspace{-13pt}
	\caption{Single-channel DP-QAM transceiver: Experimental setup (a) and DSP functions within digital coherent receiver (b).}
	\label{ExpSetup1}
	\vspace{-15pt}
	\end{figure}

	The setup of the single-channel dual-polarization (DP) QAM optical transceiver used in this work is shown in Fig.~\ref{ExpSetup1}(a) and is for the most part identical to \cite{Rob2015ECOC}. The in-phase (I) and quadrature (Q) drive signals required for shaped and uniform 16QAM and 64QAM were generated offline and digitally filtered using a root raised cosine (RRC) filter with a roll-off factor of 0.01. The symbols were loaded onto a pair of field programmable gate arrays (FPGAs) and output using two DACs, each operating at 32~GSa/s (4~Sa/sym). The electrical signals were each amplified using a linear amplifier and passed through an 8$^{\text{th}}$-order Bessel low-pass filter (LPF) with a rejection ratio of more than 20~dB/GHz and a 3~dB bandwidth of 6.2~GHz. After this amplification stage, the electrical signal and the direct output of an external cavity laser (ECL) with a 1.1~kHz linewidth were passed into an IQ modulator. The optical signal was amplified with an Erbium-doped fiber amplifier (EDFA) and polarization-multiplexed to create a Nyquist shaped DP-QAM optical carrier at 8~GBd. The DP-QAM signal was passed directly into the digital coherent receiver, which had a sample rate of 160~GSa/s and an analog electrical bandwidth of 62.5~GHz. Amplified spontaneous emission (ASE) noise was added to the signal to vary the received SNR. A second ECL (1.5~kHz linewidth) was used as a local oscillator (LO).

	The blind DSP implementation is illustrated in Fig.~\ref{ExpSetup1}(b). After being sampled at 160~GSa/s in analog-to-digital conversion stage (ADC), the received signals were corrected for receiver skew imbalance and normalized to correct for the varying responsivities of the 70~GHz balanced photodiodes of the coherent optical receiver.
	Each polarization was resampled to 2~Sa/sym before matched RRC filtering. In order to equalize the signal and to undo polarization rotations, a blind 51-tap T/2-spaced radially directed equalizer (RDE) that takes into account the probabilities of the QAM moduli \cite{DomsEQ} was used. Tap-weight pre-convergence is obtained by the constant modulus algorithm (CMA) equalizer. The symbols at the output of the equalizer were down-sampled to 1~Sa/sym and the intermediate frequency was estimated and removed using a 4$^{\text{th}}$-order nonlinearity algorithm \cite{Savory}.
	The carrier phase estimation (CPE) was performed per polarization using a decision-directed phase estimation algorithm including an averaging over a 64~T-spaced sliding window to improve the estimate \cite{Pfau}. The Gram-Schmidt orthogonalization procedure (GSOP) \cite{Fatadin2008} corrected for sub-optimal phase bias in the transmitter IQ modulators that occurred over time due to temperature variations. The symbols at the output of the GSOP stage were used to calculate an AIR estimate. Circularly symmetric Gaussian statistics are assumed for the calculation, resulting in an achievable rate for a decoder operating with these statistics, as discussed in \cite[Sec.~2]{FehenbergerOE2015}.

\section{Results}\label{sec:results}

	\subsection{Back-to-back Experiments}\label{ssec:b2b}
\begin{figure}[t]
			\centering
      \includegraphics{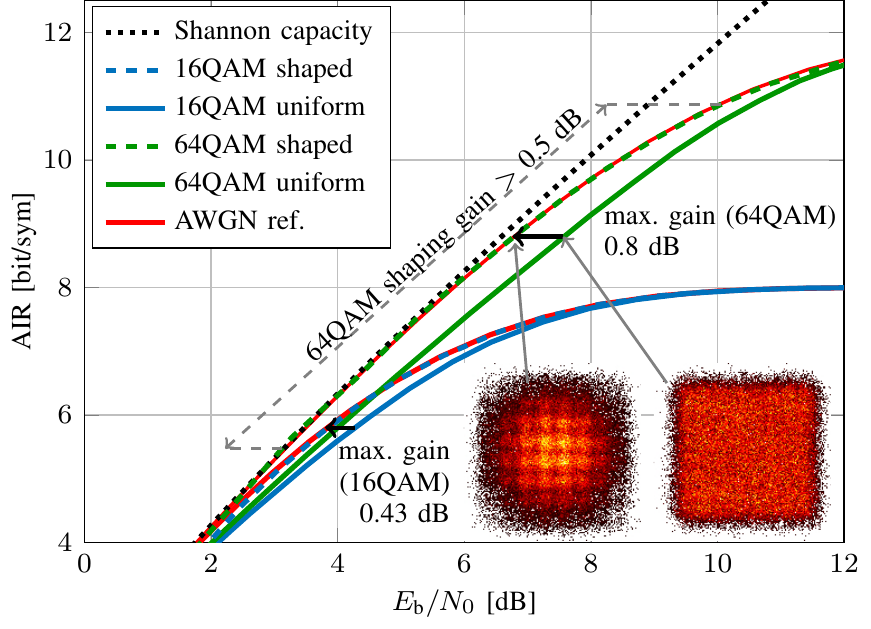}
			\setlength{\abovecaptionskip}{-15pt}
			\caption{AIRs vs. {\Eb} in B2B experiments: uniform QAM (solid lines) and probabilistically shaped QAM (dashed lines). The AWGN reference curves (red) are also shown. Insets: Received constellation diagrams for shaped (left) and uniform (right) 64QAM at the maximum shaping gain of 0.8~dB.}
			\label{fig:B2B_mQAM}
			\vspace{-5pt}
		\end{figure}

		Figure~\ref{fig:B2B_mQAM} shows for the B2B experiments AIRs vs. the energy per information bit, {\Eb}, given by $E_\text{b}/N_0={\text{SNR}}/{\text{AIR}}$, which is chosen for clarity. For both 16QAM and 64QAM, the gap to the Shannon capacity (dotted black line) is closed by shaping 16QAM (dashed blue line) and 64QAM (dashed green line). The maximum sensitivity gain for 64QAM is verified to be 0.8~dB, reducing the required {\Eb} to achieve an information rate of 8.8~bits per DP-symbol (bit/sym) from 7.6 dB to 6.8 dB. It is confirmed that 16QAM has a limited gain of at most 0.43~dB because having only 16 points restricts the shaping degrees of freedom. The gains we find are larger than the theoretical gains from geometrical shaping presented in \cite[Fig.~3]{GeoShapingExp1}, illustrating that probabilistic shaping approaches the ultimate shaping gain faster than geometrical shaping. AWGN results obtained by numerical integration are included as references (solid red line) in Fig.~\ref{fig:B2B_mQAM} and perfectly match the experimental results. For 64QAM, a significant shaping gain in excess of 0.5~dB is shown over a wide range of {\Eb} from 3.1~dB to 10~dB.

\begin{figure}[t]
			\centering
       \includegraphics{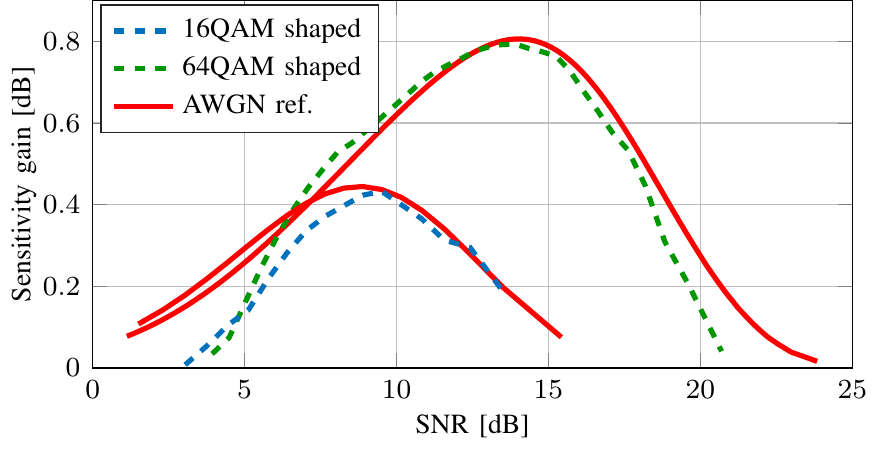}
			\setlength{\abovecaptionskip}{-15pt}
			\caption{Sensitivity gains from shaping 16QAM and 64QAM for the reference AWGN channel (solid red) and the B2B experiments (dashed).}
			\label{fig:SensitivityGain_64QAM_ExpTheory}
			\vspace{-10pt}
		\end{figure}

		In Fig.~\ref{fig:SensitivityGain_64QAM_ExpTheory}, sensitivity gains vs. SNR are depicted for 16QAM and 64QAM. A good match between theory and experiments is observed, in particular at the respective maximum sensitivity gain. The experimental curves slightly deviate from the AWGN results at low SNR due to instabilities in AIR and SNR estimation. It is interesting to note that the AWGN curves cross at approximately 7~dB SNR. The larger sensitivity gain of probabilistically shaped 16QAM in comparison to 64QAM does, however, not imply that 16QAM has a larger absolute AIR in this low-SNR range.

	 	\begin{table*}
		\caption{SNR range of the channel, input PMFs $\boldsymbol{P}_X$ and constellation points $\boldsymbol{x}$ corresponding to the arrows ($\mathrm{a}$) to ($\mathrm{d}$) in Fig.~\ref{fig:Robustness2D}.}
		\label{table:input_dists}
		\vspace{-5pt}
		\centering
		\begin{tabular}{c|c|c|c}
		Input & Channel SNR range & $\boldsymbol{P}_X$ & $\boldsymbol{x}$ \\
		\hline
			\textbf{(a)} & \hspace{10pt}5 $-$ 16.2 dB & [0.042, 0.093, 0.158, 0.207, 0.207, 0.158, 0.093, 0.042] & [-2.02, -1.44, -0.87, -0.29, 0.29, 0.87, 1.44, 2.02] \\
			\textbf{(b)} & 16.2 $-$ 19.3 dB & [0.079, 0.113, 0.145, 0.163, 0.163, 0.145, 0.113, 0.079] & [-1.73, -1.24, -0.74, -0.25, 0.25, 0.74, 1.24, 1.73] \\
			\textbf{(c)} & 19.3 $-$ 22.2 dB & [0.109, 0.122, 0.132, 0.137, 0.137, 0.132, 0.122, 0.109] & [-1.59, -1.13, -0.68, -0.23, 0.23, 0.68, 1.13, 1.59] \\
			\textbf{(d)} & \hspace{-6pt}22.2 $-$ 25 dB & [0.124, 0.125, 0.126, 0.126, 0.126, 0.126, 0.125, 0.124] & [-1.53, -1.09, -0.66, -0.22, 0.22, 0.66, 1.09, 1.53] \\
			\end{tabular}
		\vspace{-10pt}
		\end{table*}

	\subsection{Shaping Robustness}\label{ssec:robustness}

		In the experiments, the SNR for which the shaping optimization was carried out did not always exactly match the channel SNR that was measured after DSP. Strictly following the shaping optimization problem outlined in Sec.~\ref{sec:shaping}, a different PMF would be required for every SNR. The shaping gains, however, were found to be robust to a mismatch between channel SNR (measured digitally after DSP) and the shaping SNR, i.e., the SNR that was assumed at the transmitter to find the shaped input PMF.

		{To investigate this behavior in detail, we numerically calculated the shaping sensitivity gain for 64QAM over the AWGN channel. We varied both the channel SNR and the shaping SNR from 5~dB to 25~dB in steps of 0.1~dB, resulting in $201^2=40401$ combinations. We consider as a figure of merit the difference in sensitivity gain over uniform input between perfectly matched shaping, i.e., where the channel SNR equals the shaping SNR, and mismatched shaping, i.e., where the two SNRs are different.}

\begin{figure}[t]
	\vspace{-5pt}
	\centering
  \includegraphics{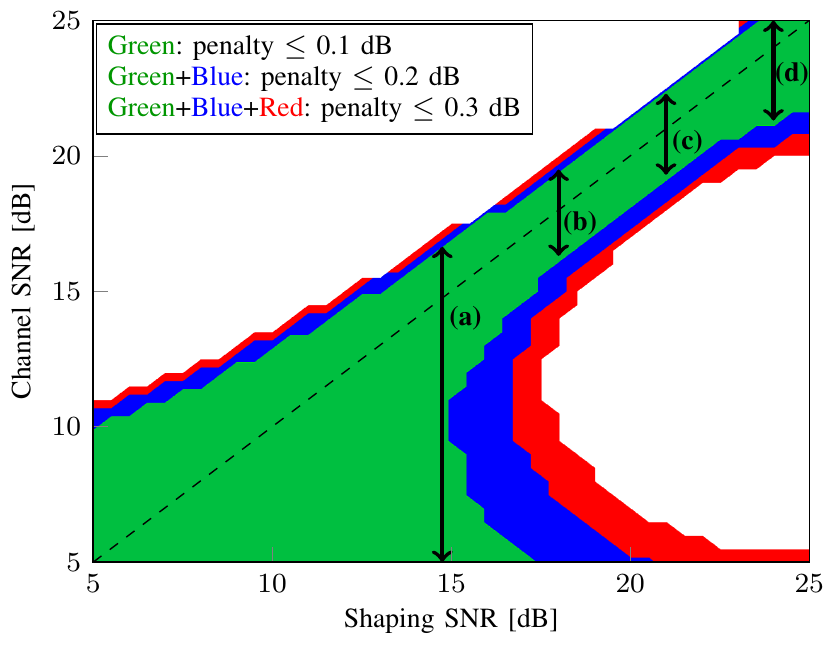}
	\setlength{\abovecaptionskip}{-5pt}
	\caption{Shaping robustness for 64QAM over the AWGN channel. Colors indicate the penalty in comparison to perfectly matched shaping. The dashed line indicates perfect shaping, i.e. 0 dB penalty, between channel SNR and shaping SNR. The arrows \textbf{(a)} to \textbf{(d)} correspond to shaping SNRs of 14.5~dB, 18~dB, 21~dB, and 24~dB, respectively.}
	\label{fig:Robustness2D}
	\vspace{-10pt}
	\end{figure}

\begin{figure}[t]
	\centering
  \includegraphics{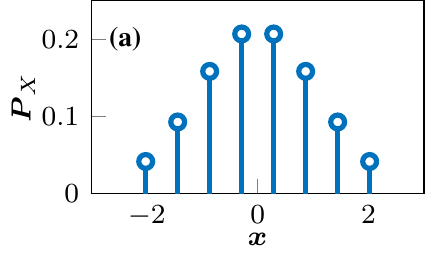}\includegraphics{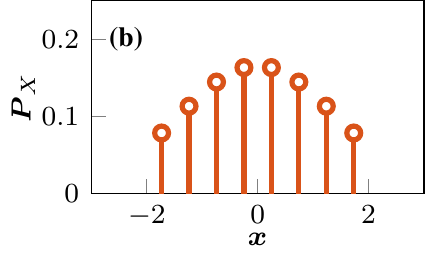}\\
  \includegraphics{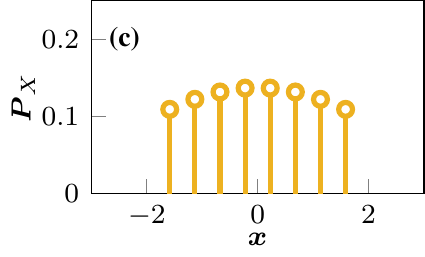}\includegraphics{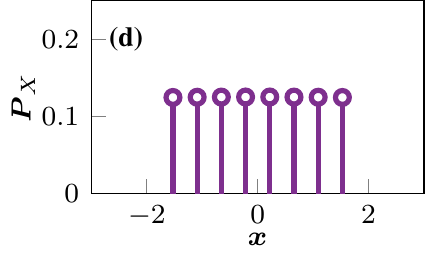}
	\setlength{\abovecaptionskip}{-5pt}
	\caption{Input PMFs for 64QAM and the four shaping SNRs \textbf{(a)} to \textbf{(d)}, which are represented as arrows in Fig.~\ref{fig:Robustness2D} and stated in the first column of Table~\ref{table:input_dists}.}
	\label{fig:input_dists}
	\vspace{-5pt}
	\end{figure}

		In Fig.~\ref{fig:Robustness2D}, the colored areas indicate all combinations of channel SNR and shaping SNR for which the largest acceptable sensitivity penalty in comparison to perfectly matched shaping (dashed black line) is below a certain threshold. The green area shows a reduction in shaping gain by at most 0.1~dB. The blue and green areas combined indicate a maximum sensitivity reduction of not more than 0.2~dB. Adding the red region increases the tolerated loss to at most 0.3~dB. We observe that in order to cover the entire relevant channel SNR range with a penalty of at most 0.1~dB SNR, only four different input PMFs are required. If a loss of 0.2~dB or 0.3~dB is acceptable, the number of PMFs is reduced to 3 or 2, respectively. The SNRs assumed for obtaining the distributions giving a 0.1~dB penalty and the covered channel SNRs are marked with black arrows in Fig.~\ref{fig:Robustness2D}. The corresponding values for the channel SNR range, the input PMF $\boldsymbol{P}_X$, and the 1D constellation points $\boldsymbol{x}$ are given in Table~\ref{table:input_dists}, providing a ready-to-use lookup-table for implementing probabilistically shaped 64QAM. Figure~\ref{fig:input_dists} illustrates how $\boldsymbol{P}_X$ starts as a Gaussian-like shape and approaches a uniform distribution as the SNR increases. Further numerical results for shaped 256QAM (not presented due to lack of space) show that four input PMFs are sufficient for a penalty of at most 0.1~dB over a SNR range from 5~dB to 30~dB.

\section{Conclusions}
	Probabilistically shaped 16QAM and 64QAM were investigated in back-to-back experiments. Sensitivity gains of up to 0.8~dB over uniform input distributions were measured experimentally, showing an excellent match with AWGN simulations {and effectively closing the gap to capacity}. {It is yet to be investigated whether there are distributions other than the one considered in this work that yield higher spectral efficiencies in nonlinear fiber transmission.}
	We further showed numerically that the shaping gain is robust to a mismatch between channel SNR and SNR assumed at the transmitter. This means that variations of the channel SNR of up to 11 dB in the low and medium SNR range, c.f. distribution a) in Table~\ref{table:input_dists}, do not require an adjustment of the input distribution, provided that a slightly reduced shaping gain is acceptable. This property makes probabilistic shaping highly suitable for a practical application in optical communication systems as it offers tolerance to SNR degradation occurring in the life cycle of an optical communication system. {Due to the Gaussianity of nonlinear interference in long-haul transmission systems, we anticipate that the conclusions would also hold in this scenario. We leave the investigation of the sensitivity of mismatched probabilistic shaping in the presence of fiber nonlinearities to future work.} 

	\section*{Acknowledgment}
	The authors would like to thank Georg B{\"o}cherer (TUM) for fruitful discussions on the shaping method described in Sec.~\ref{sec:shaping}.

	\balance 


\vspace*{-10pt}
\end{document}